\documentclass[aps,pre,twocolumn,showpacs,superscriptaddress]{revtex4-1}  

\usepackage{graphicx}   
\usepackage{dcolumn}   
\usepackage{bm}            
\usepackage{amssymb} 
\usepackage{amsmath}  

\usepackage{color}
\usepackage{subfigure}
\usepackage{braket}
\usepackage{slashed}

\def\braket#1{\mathinner{\langle{#1}\rangle}}

\newcommand{\de}{\partial}

\begin{document}

\title{Effective Distances for Epidemics Spreading on Complex Networks}
\author{Flavio Iannelli}
\thanks{F.I. and A.K. contributed equally.}
\affiliation{
Institute for Physics, Humboldt-University of Berlin, Newtonstra{\ss}e 15, 12489 Berlin, Germany
}
\email{iannelli.flavio@gmail.com}

\author{Andreas Koher}
\thanks{F.I. and A.K. contributed equally.}
\affiliation{
Institute for Theoretical Physics, Technische Universit\"at Berlin, Hardenbergstra{\ss}e 36, 10623 Berlin, Germany}

\author{Dirk Brockmann}
\affiliation{
Robert Koch-Institute, Nordufer 20, 13353 Berlin, Germany}
\affiliation{
Institute for Theoretical Biology and Integrative Research Institute of Life Sciences, Humboldt-University of Berlin,
Philippstra{\ss}e 13, Haus 4, 10115 Berlin, Germany
}

\author{Philipp H\"ovel}
\affiliation{
Institute for Theoretical Physics, Technische Universit\"at Berlin, Hardenbergstra{\ss}e 36, 10623 Berlin, Germany}

\author{Igor M. Sokolov}
\affiliation{
Institute for Physics, Humboldt-University of Berlin, Newtonstra{\ss}e 15, 12489 Berlin, Germany
}

\date{\today}

\begin{abstract}

We show that the recently introduced logarithmic metrics used to predict  disease arrival times on complex networks are approximations of more general network-based measures derived from random walks theory.  
Using the daily air-traffic transportation data we perform numerical experiments to compare the infection arrival time with this alternative metric  that is obtained by accounting for multiple walks instead of only the most probable path. The comparison with direct simulations reveals a higher correlation compared to the shortest-path approach used previously. 
In addition our method allows to connect fundamental observables in epidemic spreading with the cumulant generating function of the hitting time for a Markov chain.
Our results provide a general and computationally efficient approach using only algebraic methods. 

\end{abstract}

\maketitle

\section{Introduction}

Networks have received growing attention in the past decade particularly due to their applicability in describing a wide range of phenomena. Prominent examples are the mapping of the World Wide Web and structure of Internet, social and financial networks, epidemiology and language dynamics \cite{dynamical,internet,RevModPhys.87.925,RevModPhys.81.591}.
In the context of epidemic spreading the prediction of outbreaks has become particularly important for public health issues. 
The rapid growth in the velocity of transportation means and frequency of movements has further increased  the risk that global emergent diseases such as H1N1 \cite{Yang729}, SARS \cite{Colizza2007} or EBOV \cite{poletto2014assessing}, and more recently ZIKV \cite{Ioos2014302},  will spread worldwide.
The underlying mobility networks are usually scale-free \cite{Barabasi509}. This implies the absence of an epidemic threshold \cite{PhysRevLett.86.3200} that allows any transmittable disease to spread through the global population.

The large amount of traffic data both at the local and global scale, which became available in recent years, provides a new opportunity to understand such processes. 
On the one hand numerical simulations of infection spreading offer a practical tool for estimating the  infection arrival time \cite{Broeck2011}. 
In this regard meta-population models \cite{PhysRevE.85.066111,colizza07,colizza08}   provide a reasonable tool for maintaining a high level of complexity in the simulation of pandemics \cite{Broeck2011}.
At the global scale the different subpopulations, defined by the nodes of the network, are cities that can be connected by directed or undirected fluxes of individuals, provided by the worldwide transportation network data.
On the other hand algebraic methods give a solid foundation for drawing general conclusions and in many cases provide numerical instruments superior to direct simulations.  

\begin{figure*}
 \includegraphics[scale=0.32]{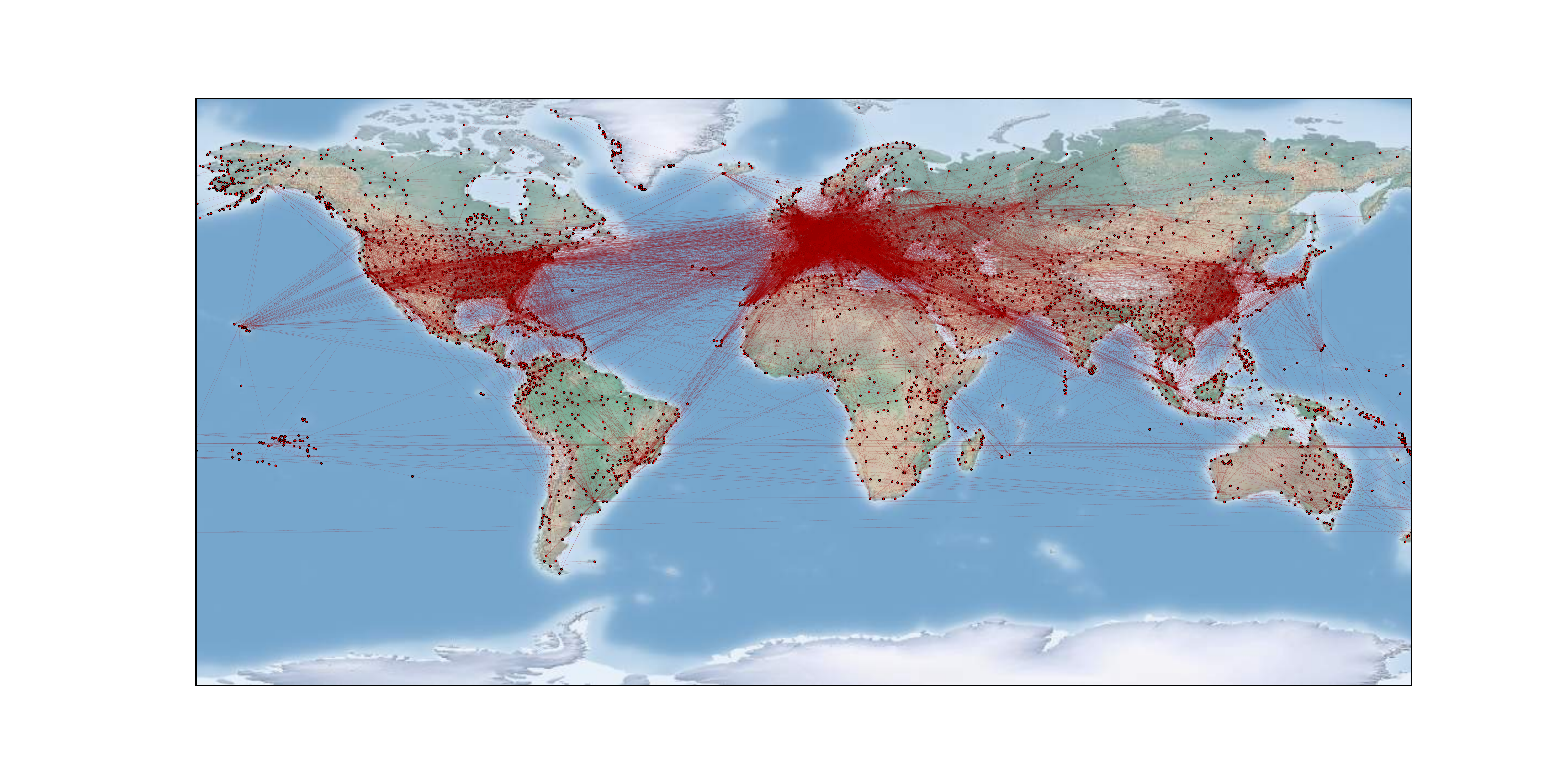}
  \caption{The global mobility network used in the simulations consisting of $V = 3865$ airports and $E = 51440$ flights \cite{data}.}
  \label{graph}
\end{figure*}

In this work we introduce a network-based measure that generalizes the concept of distance and that provides fundamental insights into the dynamics of disease transmission as well as a efficent numerical estimation of the infection arrival time. We compare this \emph{effective distance} \cite{bara_netsci} with the numerical estimate of the transmission times using a meta-population model to validate the method.
A series of papers have already been devoted to this problem \cite{PhysRevLett.91.168701,gautreau2007arrival,gautreau2008global,brockmann2013hidden,Lawyer2016}. Most of them  rely on the concept of most probable path, the shortest-path effective distance $D^{\text{SP}}_{ij}$ for each source $i$ and target $j$ in the network. 
The latter can be defined, for both directed and undirected networks, as the geodesic graph distance with  edge weights given by the first moment of a Gumbel distributed variable which depends only on the network topology and the infection rate. This shortest-path approach however significantly overestimates the infection arrival times \cite{gautreau2007arrival,crepey2006epidemic}. A more realistic scenario takes into account all possible paths \cite{gautreau2008global} yielding the multiple-path distance $D^{\text{MP}}_{ij}$ which is better suited to estimate the arrival times of the infection. 
This method allows, at least in principle, to take into account all possible directed transmission paths although the computation becomes infeasible as their number grows exponentially with the the number of vertices in the network. The lack of a practical computational approach leads back to considering only the shortest path.
A logarithmic ad-hoc edge weight transformation was introduced in \cite{brockmann2013hidden}  by simply requiring that adding edges should translate to multiplying the associated probabilities. This follows the intuitive argument that a higher number of passengers reduces the separation distance between neighbouring nodes. 
This logarithmic  transformation can be viewed as a log-space reduction \cite{roosta1982routing} and, how we will show later in the text, it can be derived as a special case of the shortest-path effective distance defined previously in \cite{gautreau2007arrival}.
The estimated arrival time $T^{\text{AR}}_{ij}$ from node $i$ to node $j$, obtained from numerical simulations, correlates highly with the shortest-path effective distance. Using this metrics the complexity of disease transmission can be understood in terms of circular waves propagating at constant velocity from the infected node at time zero to all other nodes in the network \cite{brockmann2013hidden}.
The measure we introduce here aims to generalize previous definitions by including all walks that connect source and target.

\section{Epidemiological model}

\begin{figure}
 \subfigure[]{\includegraphics[scale=0.38]{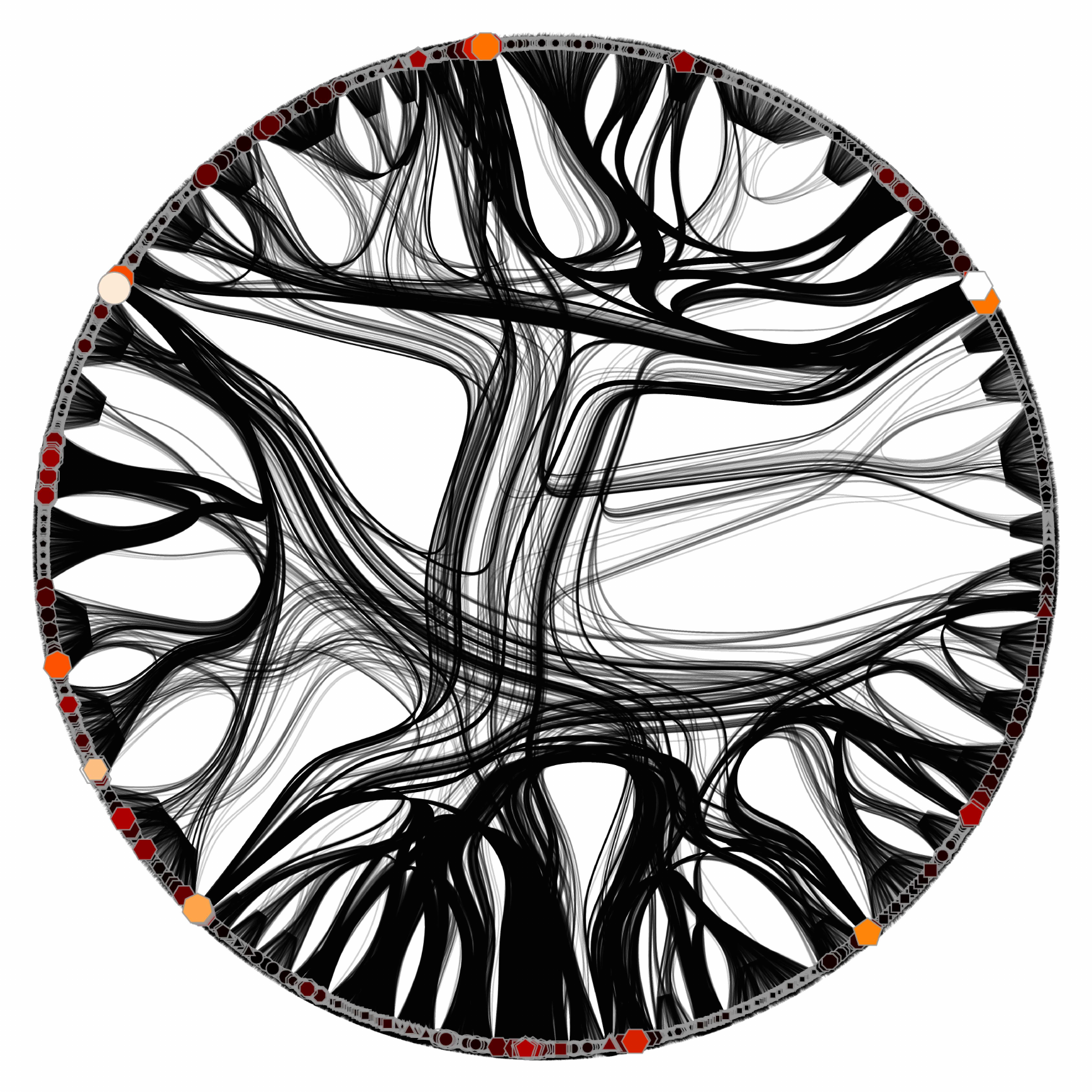}}
 \subfigure[]{\includegraphics[scale=0.42]{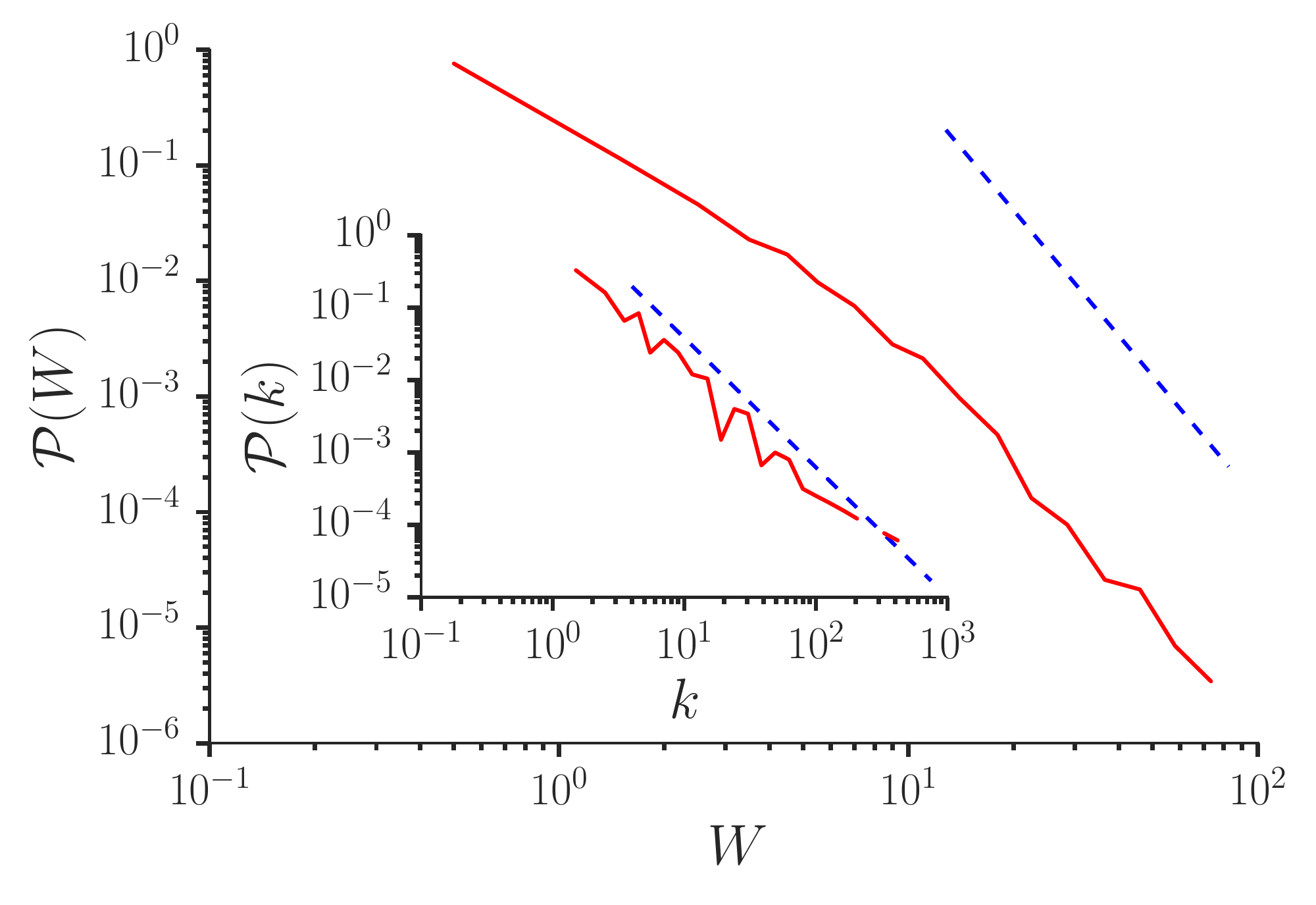}}
  \caption{\label{powerlaw}(Color online) (a) Circular representation of the global mobility network with vertex colors and size corresponding to its strenght $k_i^W$. (b) Normalized weight distribution $\mathcal{P}(W) \sim W^{-\gamma_w}$ with scaling exponent $\gamma_w = 3.60 \pm 0.14$. Inset: unweighted degree distribution $\mathcal{P}(k) \sim k^{-\gamma_k}$ with scaling exponent $\gamma_k = 1.79 \pm 0.10$. Scaling exponents are obtained using the method presented in \cite{doi:10.1137/070710111}.}
\end{figure}

Let us consider a meta-population susceptible-infected-removed  dynamics $S\xrightarrow{\beta SI}I \xrightarrow{\mu I} R$, where $\beta$ and $\mu$ are the infection and recovery rate, respectively. The nodes of the meta-population network consists of subpopulations of constant size $N_j$, which divides into susceptible ($S_j$), infected ($I_j$) and removed ($R_j$) individuals
\begin{equation}
N_j = S^{(t)}_j + I^{(t)}_j +R^{(t)}_j .
\label{norm}
\end{equation}
In the meta-population in addition to the local SIR reaction dynamics, the movement of a host between populations $i$ and $j$ is governed  by the master equation
\begin{equation}
\de_t X_j^{(t)} =  \sum_{i\ne j} \left(X^{(t)}_i Q_{ij}  - X^{(t)}_j Q_{ji}\right).
\label{kineticbrock2} 
\end{equation}
Here, we introduced $X^{(t)}_j=\{S^{(t)}_j, I^{(t)}_j, R^{(t)}_j\}$ as a placeholder variable and the transition rates $Q_{ij}$ are defined as the conditional probability of a randomly chosen individual to jump from location $i$ to location $j$ within one time step
\begin{equation}
\mathbb{P}\left(X^{(t+\Delta t)}_j \middle|  X^{(t)}_i\right) \approx Q_{ij}\Delta t, \qquad i\ne j.
\label{jump_prob}
\end{equation}
The transition rates $Q_{ij}$ can equivalently be defined in terms of the weighted adjacency matrix $A^W_{ij}$ as $Q_{ij} = {A^W_{ij}}/{N_i} \in [0,1]$. The latter is obtained from the actual passenger fluxes  between any two airports  in the global mobility network used in the simulations \cite{data}. This network consists of $V = 3865$ vertices (airports) and $E = 51440$ directed edges (fluxes), with very broad degree and weight distributions, see Fig.~\ref{graph} and ~\ref{powerlaw}, where the high heterogeneity in the network connectivity is graphically reproduced. 
For the network diameter we found $D=16$ (connecting Stuart Island to Narsaq Kujalleq Heliport) and the global clustering coefficient is  $c = 0.26 \pm 0.01$.
A peculiar feature of this network is that the antisymmetric part of the fluxes $\chi = A^W_{ij} - A^W_{ji}$  is vanishing to a high degree of accuracy so that it can be considered as undirected \cite{barrat2004architecture}. The weighted adjacency matrix  of the undirected air traffic network is then defined by $A^W_{ij} = A_{ij}W_{ij}$, where $A_{ij}$ is the adjacency matrix element and $W_{ij}\ge 0$ the corresponding weight. 
The symmetry in  the adjacency matrix implies that for adjacent populations 
\begin{equation}
A^W_{ij} = Q_{ij} N_i = Q_{ji} N_j = A^W_{ji}.
\label{detbal}
\end{equation}
The Markov transition matrix associated to the network can be written in terms of both the fluxes $ A^W_{ij}$ and the local transition rates $Q_{ij}$
\begin{equation}
P_{ij} = \frac{A^W_{ij}}{\sum_j A^W_{ij}}  = \frac{Q_{ij}}{\sum_j Q_{ij}}.
\label{normal_brock}
\end{equation}
and it is row stochastic by construction.
From \eqref{detbal} we also have detailed balance 
\begin{equation}
P_{ij} k^W_i = P_{ji} k^W_j,
\label{nice}
\end{equation}
where we have introduced the weighted degree $k^W_i = \sum_j A^W_{ij}$, sometimes denoted as strength \cite{calda}, as the  asymptotic probability distribution for the corresponding Markov chain. 
Thus using \eqref{nice} we can rewrite \eqref{kineticbrock2}, in terms of the compartment densities $x_j^{(t)} = X_j^{(t)}/N_j$ to obtain
\begin{equation}
\begin{aligned}
\de_t x_j^{(t)} 
& = 
\frac{1}{N_j} \sum_{i\ne j} \left(x^{(t)}_i A^W_{ij} - x^{(t)}_j A^W_{ji}\right) \\
& =
\frac{k^W_j}{N_j} \sum_{i\ne j} P_{ji}   \left(  x^{(t)}_i - x^{(t)}_j \right).
\end{aligned}
\label{kineticbrock3} 
\end{equation}
Furthermore, we can remove the dependence on the population size $N_j$ by introducing a constant global mobility rate $\alpha$. This parameter is defined as the ratio $\alpha = {\Phi}/{N}$, between the total daily passenger flux $\Phi = \sum_{ij} A_{ij}^W$ and the total population $N=\sum_i N_i$, i.e. the rate to leave a node for a randomly  chosen individual. 
A local node dependent mobility rate can also be defined as 
\begin{equation}
\alpha_i=\frac{\sum_j A^W_{ij}}{N_i} = \sum_j Q_{ij}.
\label{alphai}
\end{equation}
In the global mobility network data  the total traffic  of each  node, i.e. its weighted degree $k^W_j$, is with a good accuracy proportional to its population $N_i$ via the global mobility rate $\alpha$, thus in our case $\alpha_i = \alpha$ $\forall i$. The complete SIR dynamics of the Rvachev-Longini model \cite{rvachev1985mathematical,brockmann2013hidden} becomes
\begin{equation}
\begin{cases}
\de_t s_j^{(t)}  =   \Omega(\{s_j^{(t)}\})  -\beta s_j^{(t)} i_j^{(t)}\\
\\
\de_t i_j^{(t)} =  \Omega(\{i_j^{(t)}\})  +\beta s_j^{(t)} i_j^{(t)} - \mu i_j^{(t)} \\
\\
\de_t r_j^{(t)} = \Omega(\{r_j^{(t)}\})   + \mu i_j^{(t)} 
\end{cases} 
\label{system}
\end{equation}
where the mobility function for each compartment density $x_j = X_j/N_j$
\begin{equation}
\Omega(\{x_j\}) =  \alpha \sum_{i\ne j} P_{ji}   \left(  x_i^{(t)} - x_j^{(t)} \right),
\end{equation}
accounts for diffusion along the nodes.

Integrating system \eqref{system} we obtain the contagion dynamics on the transportation network $A^W_{ij}$ with the global mobility rate $\alpha$ and the infection parameters $\beta$ and $\mu$.
Finally, the arrival time ${T}^{\text{AR}}_{ij}$ for each source-target  pair in the global mobility network is computed by considering a node $j$ infected as soon as one infected individual is present.
After introducing the effective distance we use ${T}^{\text{AR}}_{ij}$ to compare the goodness of the different measures.

\section{Effective network-based measures}

The fundamental metric on a network is given by the (geodesic) shortest-path length over all paths $\Gamma_{ij}$ connecting node  $i$ to node  $j$. In a weighted network for each edge $(k,l) \in \Gamma_{ij}$ no node is visited more than once and contributes to the total length with its corresponding weight \cite{dynamical}
\begin{equation}
D_{ij} = \min_{\{\Gamma_{ij}\}}  \sum_{\substack{(k,l)\in \Gamma_{ij}}} \frac{1}{A^W_{kl}},
\label{dgeo}  
\end{equation}
where the inverse is used because in our case a higher flux of passengers reduces the distance between two nodes.
Starting from this definition it is possible to extend the notion of distance by replacing the adjacency matrix weight with an effective quantity that captures the essence of the problem of predicting when the disease will arrive at a certain place.
In \cite{gautreau2007arrival}  the authors defined the shortest-path distance $D^{\text{SP}}_{ij}$ by first considering the susceptible-infected model with only two cities, by then generalizing to arbitrary topologies. 
We first show how their analytical approach leads to an equivalent definition of effective distance used in \cite{brockmann2013hidden} and then we generalize it to more realistic propagation scenarios where all paths are taken into account. 

Let us consider two susceptible populations $i$ and $j$ and place an infected individual in $i$ at time $t_i = 0$. Assuming $Q_{ij} \Delta t \ll 1$ we can derive a probability density function for the infection hitting time $h_j$ to city $j$ of the  Gumbel type \cite{gautreau2007arrival}.
The first moment of this distribution is given by
\begin{equation}
\braket{h_j}_i =  \frac{1}{\beta}\left(\ln \frac{\beta }{Q_{ij}} - \gamma_e  \right),
\end{equation}
where $\gamma_e  \approx 0.5772$ is the Euler-Mascheroni constant and the average $\braket{\dots}_i$ is taken over times starting at $t_i = 0$.
Using \eqref{normal_brock} we find 
\begin{equation}
\begin{aligned}
\beta \braket{h_j}_i 
 & =
 \ln \frac{\beta}{\sum_j Q_{ij}}  - \gamma_e   - \ln \frac{Q_{ij}}{\sum_j Q_{ij}}\\
 & =
\delta - \ln P_{ij}  ,
\end{aligned}
\end{equation}
where we assume the mobility rate \eqref{alphai} to be  node independent, i.e. $\alpha_i = \alpha$ $\forall i$ and $\delta =  \ln (\beta/\alpha) - \gamma_e$ is a dimensionless parameter. 
This result can easily be generalized to the SIR model and a network with an arbitrary number of nodes simply by minimizing the quantity $\braket{h_l}_k$ for each consecutive link $(k,l)$ that belongs to the  path $\Gamma_{ij}$ connecting source $i$ to target $j$. 
For the arbitrary heterogenous population and network topology with an arbitrary number of nodes by taking the minimum over all paths yields the shortest-path effective distance
\begin{equation}
D^{\text{SP}}_{ij}(\delta) \equiv \min_{\{\Gamma_{ij}\}}  \sum_{\substack{(k,l)\in \Gamma_{ij}}} \left(\delta  -   \ln P_{kl}\right) \approx  (\beta - \mu) \braket{h_j}_i,
\label{effchi}  
\end{equation}
where for the SIR meta-population dynamics we have 
\begin{equation}
\delta =\ln \left(\frac{\beta - \mu}{\alpha}\right) - \gamma_e.
\end{equation}
Since each term in the sum is strictly positive, one can obtain the complete shortest-path distance matrix for each source and target pair  using  the Dijkstra algorithm  \cite{dijkstra1959note} in a time $\mathcal{O}(VE + V^2\log V)$, where $E$ and $V$ are the graph size and order, respectively. 
The effective distance defined in \cite{brockmann2013hidden} can then be recovered as a special case of \eqref{effchi} simply by setting the scale parameter $\delta$ to unity. This fact allows for a deeper and more complete understanding of this effective distance and offers a more solid  explanation  to the extremely high correlation with the infection arrival time found in \cite{brockmann2013hidden}.
The most important limitation of \eqref{effchi} is that only one path is considered, namely  the path that in addition to minimizing the topological length also maximizes the probability associated to that. Thus in this scenario the contribution to the disease spread comes only from a single route. 
The effective infection arrival time $D^{\text{SP}}_{ij}(\delta)/\mathcal{V}^{\text{EF}}$, where $\mathcal{V}^{\text{EF}} \approx \beta-\mu$ is the linearized effective speed of the infection \cite{gautreau2007arrival,brockmann2013hidden}, is in fact  overestimated  with respect to the simulations result $T^{\text{AR}}_{ij}$ \cite{crepey2006epidemic}.

\begin{figure*}
\centering
 \includegraphics[width=0.7\textwidth]{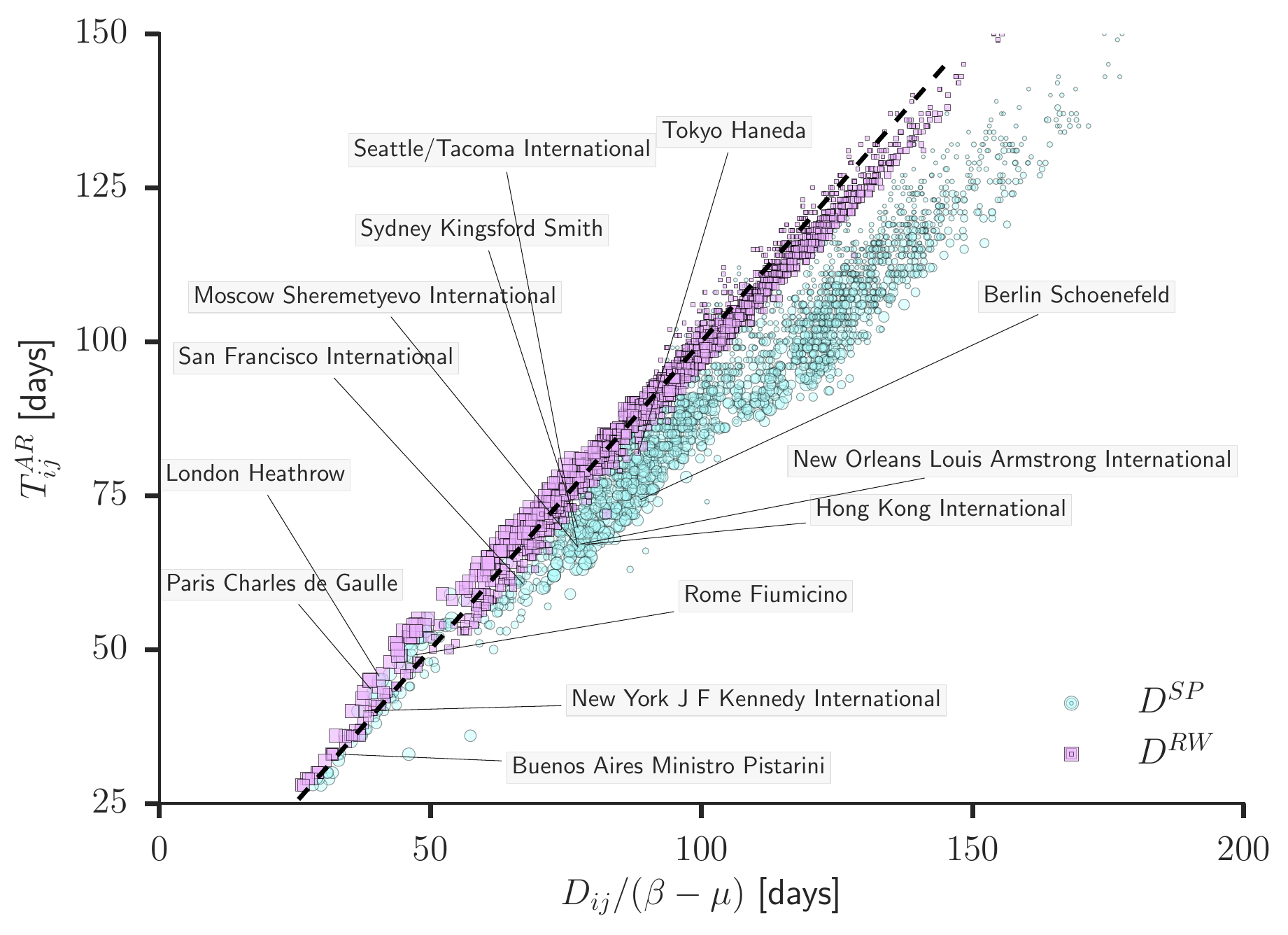}
  \caption{(Color online) Correlation of the infection arrival times in days obtained from simulation of \eqref{system}  with the shortest-path (light-blue circles) and the random-walk effective distance (violet squares). The source infected node $i$ is  S\~ao Paulo Guarulhos International  Airport and each point in the scatter plot  corresponds to an airport $j$ in the global mobility network, with size proportional to its strength $k^W_j$. The mobility and epidemiological parameters are respectively $\alpha = 0.028$ d$^{-1}$, $\beta =  0.407$  d$^{-1}$ and  $\mu = 0.271$  d$^{-1}$ resulting in $\delta = 1$. The Pearson correlation coefficients are $R^2_{\text{SP}} = 0.96$ and $R^2_{\text{RW}} = 0.99$.}
  \label{corr1}
\end{figure*}

To take into account the most realistic disease spread scenario one has to consider instead the multiplicity of transmission routes. 
For two distinct paths $\Gamma_{ij}$ and $\Gamma_{ij}'$ connecting $i$ with $j$, the authors in \cite{gautreau2008global} found that a two path distance $D^{\text{2P}}_{ij}$ satisfies the equation
\begin{equation}
e^{-D^{\text{2P}}_{ij}} =  e^{-D^{\Gamma_{ij}}} + e^{-D^{\Gamma_{ij}'}},
\label{twopaths}
\end{equation}
where 
\begin{equation}
D^{\Gamma_{ij}}(\delta) = \ln \left(\prod_{(k,l) \in \Gamma_{ij}} \frac{e^{\delta}}{ P_{kl} }\right)
\end{equation}
is the effective distance associated to the path $\Gamma_{ij}$ of arbitrary length, which corresponds to a Gumbel distributed hitting time. Relation \eqref{twopaths} can then be easily generalized to an arbitrary number of multiple paths as
\begin{equation}
\exp\left(-D^{\text{MP}}_{ij}(\delta)\right) = \sum_{\{\Gamma_{ij}\}} \exp\left(-D^{\Gamma_{ij}}(\delta)\right),
\end{equation}
so that we obtain 
\begin{equation}
\begin{aligned}
D^{\text{MP}}_{ij}(\delta) 
& =
-\ln \left(\sum_{\{\Gamma_{ij}\}}  e^{-n_{ij}\delta } F_{ij}(\Gamma_{ij})  \right).
\label{effchi1}
\end{aligned}
\end{equation}
Here we have defined the total  probability associated to the path $\Gamma_{ij}$  as
\begin{equation}
F_{ij}(\Gamma_{ij}) = \prod_{(k,l) \in \Gamma_{ij}} P_{kl}.
\end{equation}
By grouping all probabilities associated to paths of the same length in the quantity $F_{ij}(n) = \sum_{|\Gamma_{ij}| = n} F_{ij}(\Gamma_{ij})$, we can replace in \eqref{effchi1}  the sum over all paths connecting $i$ to $j$ with a sum over integer path lengths to get
\begin{equation}
D^{\text{MP}}_{ij}(\delta) = -\ln \left(\sum_{n=1}^{n_{max}} e^{-n\delta } F_{ij}(n)  \right),
\label{effchi2}
\end{equation}
where $n_{max}$ is the maximum path length in the network.
If we select the path $\widetilde{\Gamma_{ij}}$ of length $\widetilde{n}$ that is associated to the dominant contribution, i.e. the path that maximize its associated probability and minimizes the topological path length, one recovers the shortest-path effective distance of \eqref{effchi} 
\begin{equation}
\widetilde{D^{\text{MP}}_{ij}} (\delta) = \widetilde{n}\delta  - \ln F_{ij}(\widetilde{n}) = D^{SP}_{ij}(\delta).
\label{guiz}
\end{equation}
Therefore the multiple-path distance gives a more accurate estimate of the infection arrival time, as it allows to take into account the most probable route as well as all possible alternative  transmission routes.
However since the total number of paths between $i$ and $j$ can scale as $\mathcal{O}(V!)$, the measure $D^{\text{MP}}_{ij}$ becomes  computationally infeasible for large graphs. 

Both measures  ${D}^{\text{SP}}_{ij}$ and $D^{\text{MP}}_{ij}$ rely on the fact that the epidemic will spread along simple paths, i.e. routes that do not cross themselves.
Here we follow instead a different approach and introduce a distance that includes all possible random walks from source to target. Relaxing the assumption of directed spread is equivalent to effectively erasing the memory from the system at each time step. This is achieved by including in \eqref{effchi1} all walks $\Xi_{ij}$, which contrary to the paths $\Gamma_{ij}$,  allow also already visited nodes. 
We define the random-walk effective distance by generalizing \eqref{effchi1}  as 
\begin{equation}
D^{\text{RW}}_{ij}(\delta) = -\ln \left(\sum_{\{\Xi_{ij}\}}  e^{-n_{ij}\delta } H_{ij}(\Xi_{ij})  \right).
\label{effrw1}
\end{equation}
where $H_{ij}(\Xi_{ij})$ is the probability associated to a walk  that starts in $i$ and arrives to $j$. As for the probabilities $F_{ij}$ we can group the probabilities associated to walks of the same length into $H_{ij}(n) = \sum_{|\Xi_{ij}| = n} H_{ij}(\Xi_{ij})$. The latter is precisely the hitting time probability for a Markov chain defined recursively as \cite{norris1998markov}
\begin{equation}
H_{ij}(n) = \sum_{k\ne j} P_{ik} H_{kj}(n) .
\end{equation}
Thus $H_{ij}(n)$ is simply the $n$-th power of the sub-transition probability matrix obtained by removing  the $j$th row and column. 
Contrary to the multiple-path scenario now the walks are unbounded and so becomes  $n_{max}$.  Furthermore since each term in the sum \eqref{effrw1} is positive, assuming the convergence of the sum, we can rearrange it as
\begin{equation}
D^{\text{RW}}_{ij}(\delta) = -\ln \left(\sum_{n=1}^\infty e^{-n\delta} H_{ij}(n)\right).
\label{rweff}
\end{equation}

\begin{figure}
\centering
  \includegraphics[width=0.45\textwidth]{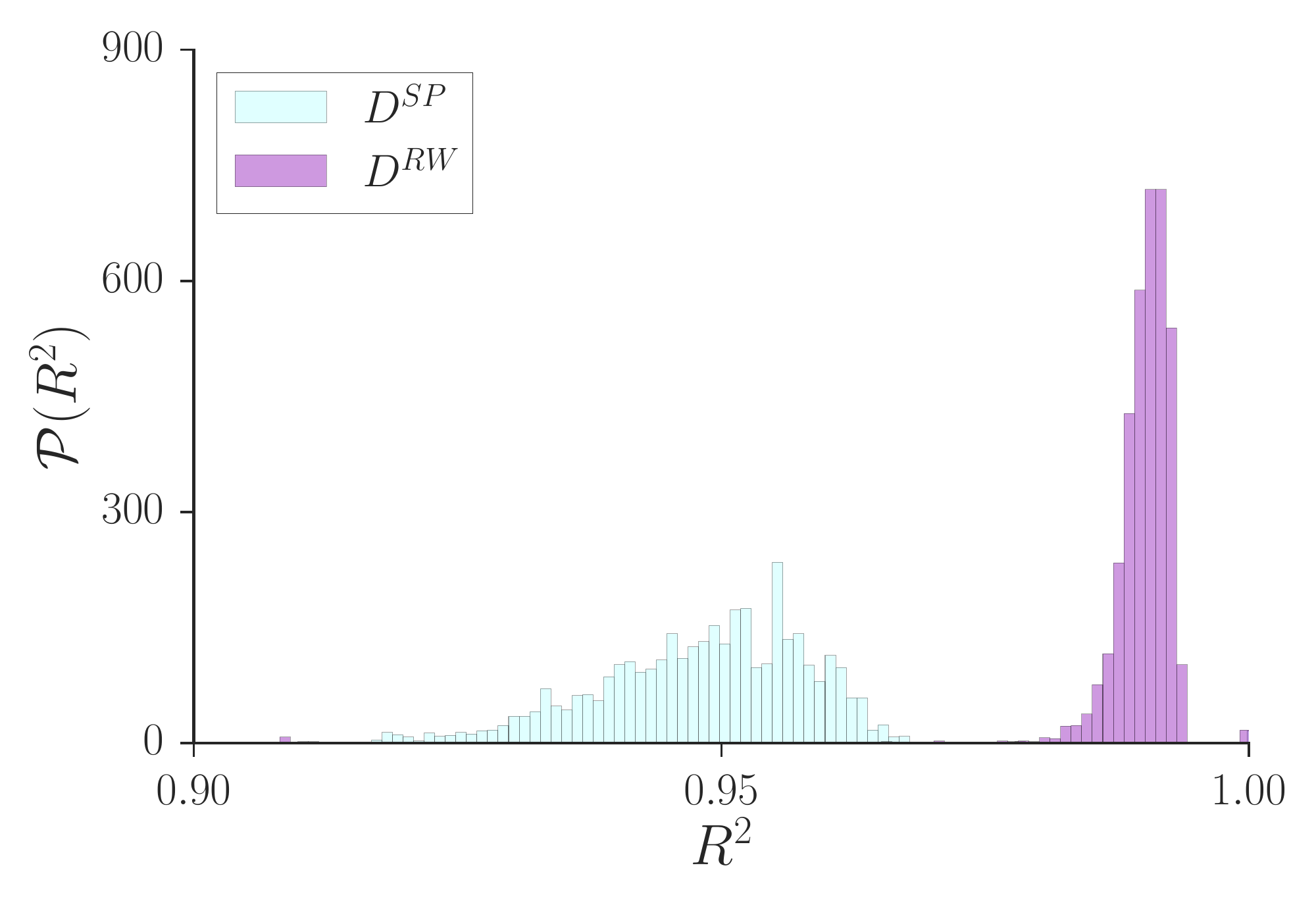}
  \caption{(Color online) Distribution of the Pearson coefficient $R^2$ considering all possible infection sources in the global mobility network for same parameters as in Fig.~\ref{corr1}.}
  \label{corr2}
\end{figure}

In Fig.~\ref{corr1} we use the Pearson correlation coefficient $R^2$ for quantifying the accuracy of the different measures using S\~ao Paulo airport as the source of the infection. Each dot in the scatter plot corresponds to an airport, which is labelled infected in the simulations when the the infection density is greater than zero. The high correlation with the infection arrival time found in \cite{brockmann2013hidden} using a shortest-path approach (light-blue) is improved when considering the random-walk effective distance (violet). The points on the dashed diagonal indicate a perfect agreement between the simulation and the effective distance.  
The correlation distribution considering all nodes in the network as initial infected seed  shows that not only the measure proposed here possesses a higher averaged correlation but it is also more peaked around it, see Fig.~\ref{corr2}.

\begin{figure*}[]
\subfigure[]{\includegraphics[scale=0.28]{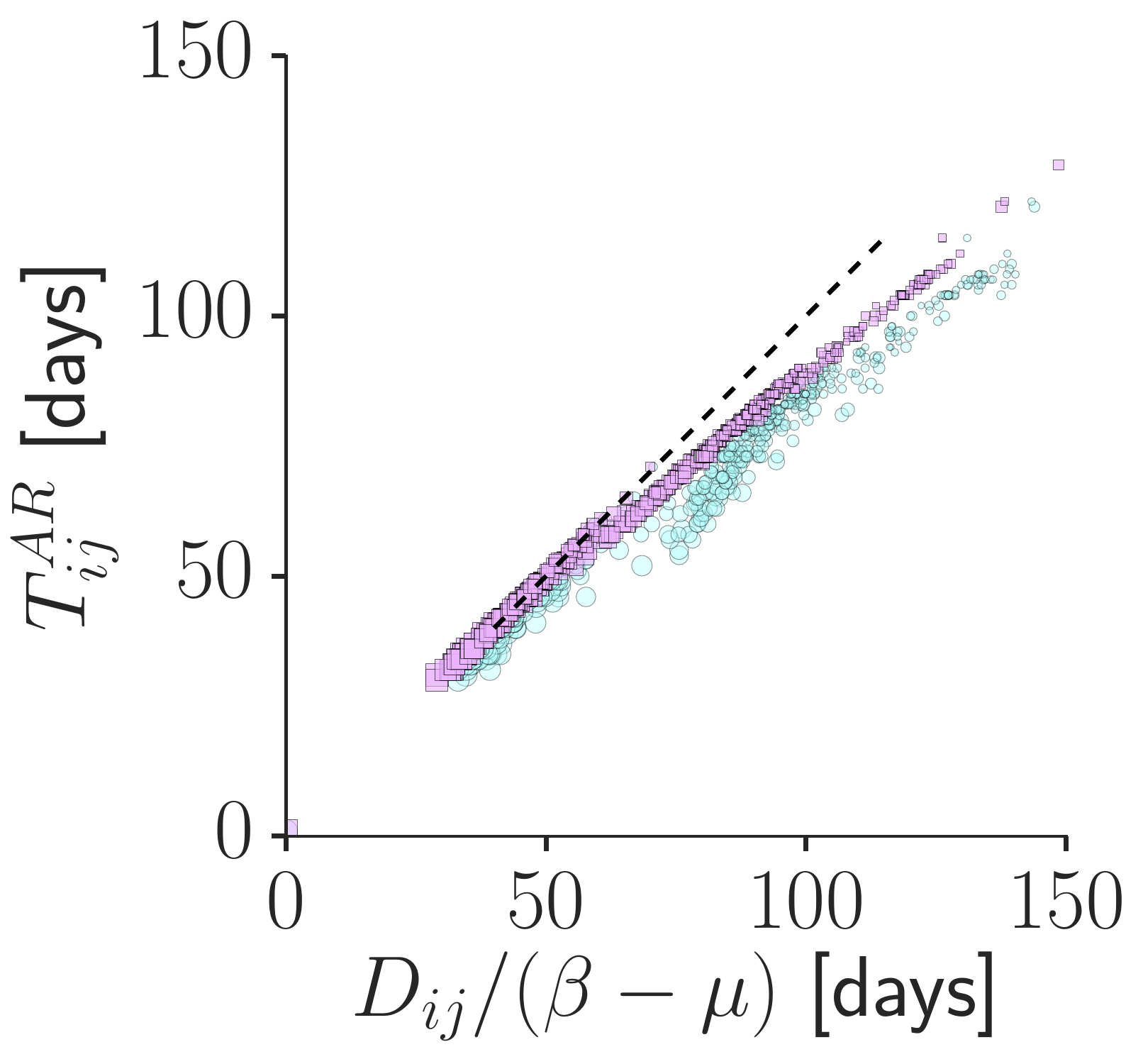}}
\qquad
\subfigure[]{\includegraphics[scale=0.18]{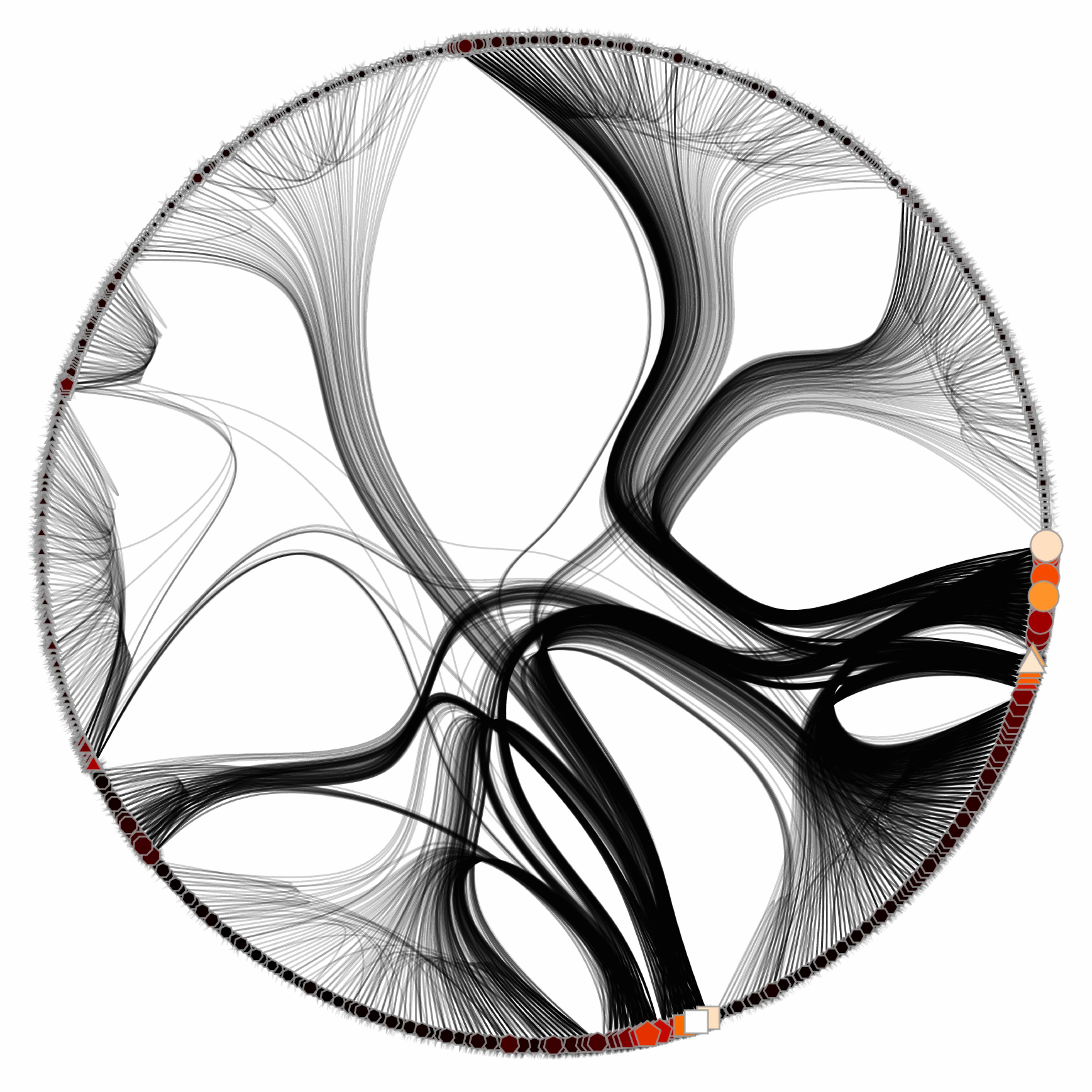}}
\qquad
\subfigure[]{\includegraphics[scale=0.28]{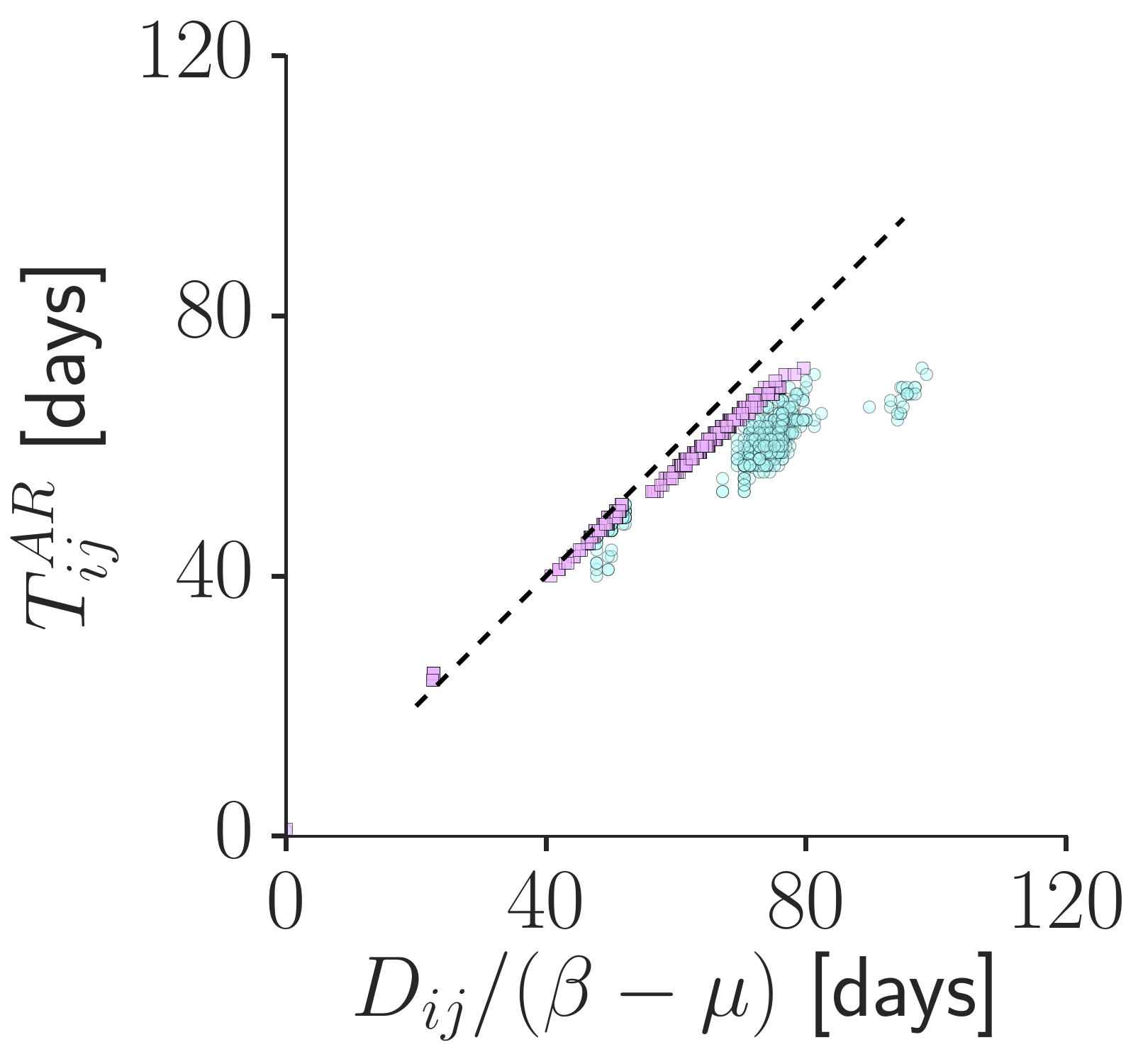}}
\qquad
\subfigure[]{\includegraphics[scale=0.18]{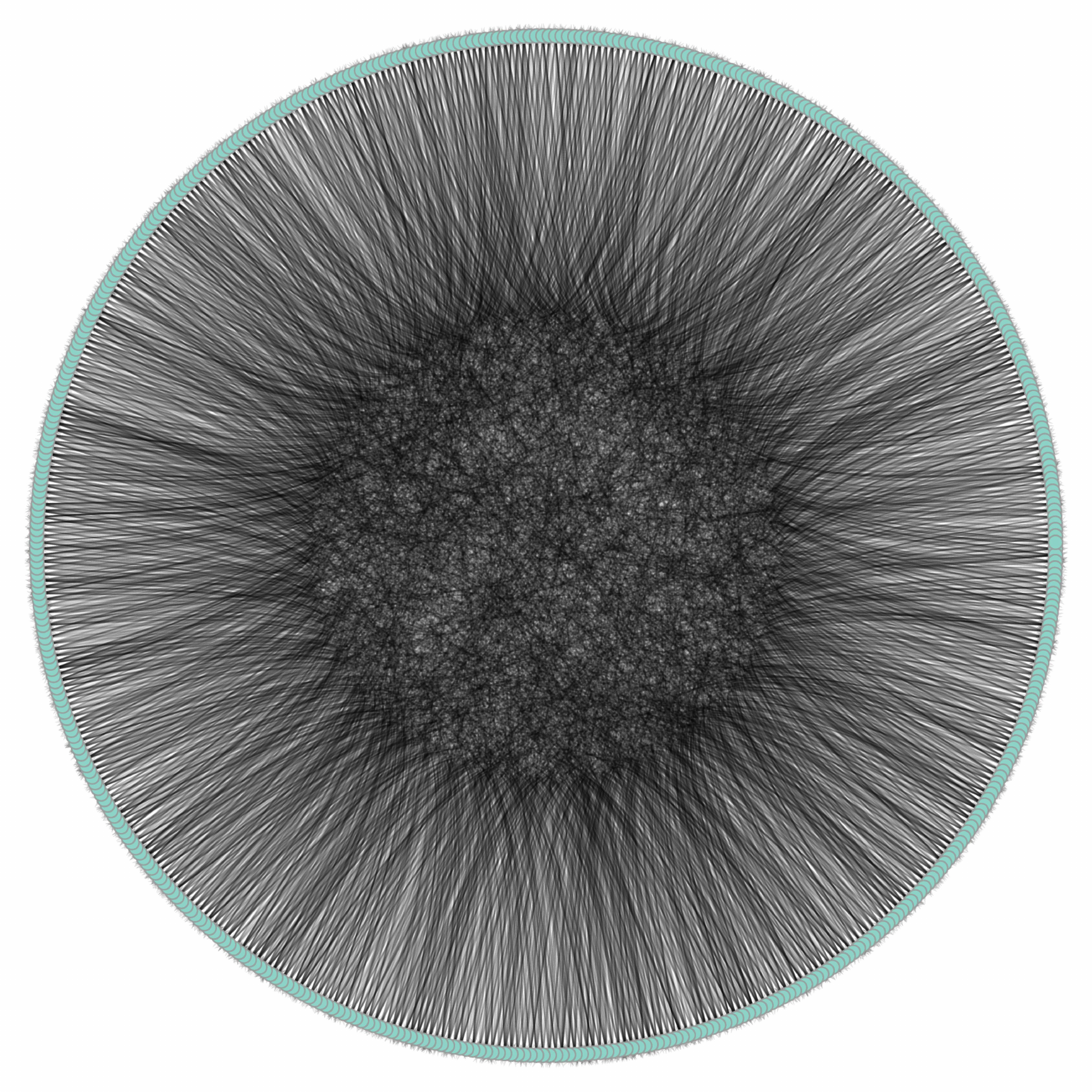}}
\caption{\label{corr_usa} (Color online) Correlation of the shortest-path (light-blue)  and random-walk (violet) approach with the simulations arrival time for the the USA airport network used in \cite{colizza07} (a)  and its randomized (Erd\H{o}s-R\'{e}nyi)  version  (c). In (b) and (d) the corresponding networks visualisation consisting of $V=500$ nodes and $E=5960$ edges. Parameters as in Fig~\ref{corr1}.} 
\end{figure*}

A remarkable interpretation of  the random-walk effective distance can be found  by noticing that by definition  $D^{\text{RW}}_{ij} (\delta) = - \ln \braket{e^{-{\delta h_j}}}_i$, where $h_j$ is the hitting time to node $j$ \cite{kemeny1960finite}. Thus, since $H_{ij}(0)=0$ for $i\ne j$, we have the correspondence 
\begin{eqnarray}
D^{\text{RW}}_{ij} (\delta) = -C_{ij}(-\delta),
 \end{eqnarray}
with the logarithm of the moment generating function, i.e. the cumulant generating function of the hitting time in a Markov chain 
\begin{equation}
C_{ij}(s) = \ln \left(\sum_{n=0}^\infty e^{ns}  \ H_{ij}(n) \right) = \ln \braket{e^{sh_j}}_i.
\end{equation} 
Hence one obtains the cumulants of the hitting time by differentiating the random-walk effective distance with respect to $\delta$ at $\delta = 0$. This interesting correspondence allows one  to rigorously relate  epidemiological quantities such as the arrival time and the speed of infection in a reaction-diffusion model  to the fluctuations of the hitting time. Then one can interpret $D^{\text{RW}}_{ij} (\delta)$ as a generalized free energy in a statistical physics perspective \cite{Kivimaki2014600} and providing a more profound theoretical framework than the ad hoc measure proposed in \cite{brockmann2013hidden}. 

From the computational side, in order to evaluate the infinite sum in \eqref{rweff}, we can restrict ourselves to the first non-vanishing contributions, which dominate due to the decreasing exponential in the walk length $n$.
However, we can also solve the complete expression by rewriting \eqref{rweff} into a geometric series.
This requires to vectorize $D^{\text{RW}}_{ij}$ with respect to the arrival node $j$ to obtain
\begin{equation}
d^{\text{RW}}_{i{(j)}} (\delta)
 =
  -\ln  \left[ \left(e^\delta \mathbf{I}(j|j)- \mathbf{P}(j|j)\right)^{-1} \mathbf{p}(j) \right]_i,
\end{equation}
where $\mathbf{P}(j|j)$  and $\mathbf{I}(j|j)$ are  the transition and identity submatrices obtained by deleting row and column $j$ \cite{zhang2011matrix},  while $\mathbf{p}(j)$ is the $j$-th column of $\mathbf{P}$ with element $j$ removed. To obtain the previous expression we have used that for $\delta > 0$, all eigenvalues of the matrix $e^{-\delta} \mathbf{P}(j|j)$ are strictly smaller than unity. 
For each arrival node the random-walk effective distance can then be obtained in polynomial time $\mathcal{O}(V^{3.4})$ using for instance the Coppersmith-Winograd algorithm for matrix inversion \cite{coppersmith1987matrix}, making the problem of parallel transmission routes feasible even for large networks as the one used in our simulations.

\begin{figure*}[]
\subfigure[]{\includegraphics[scale=0.28]{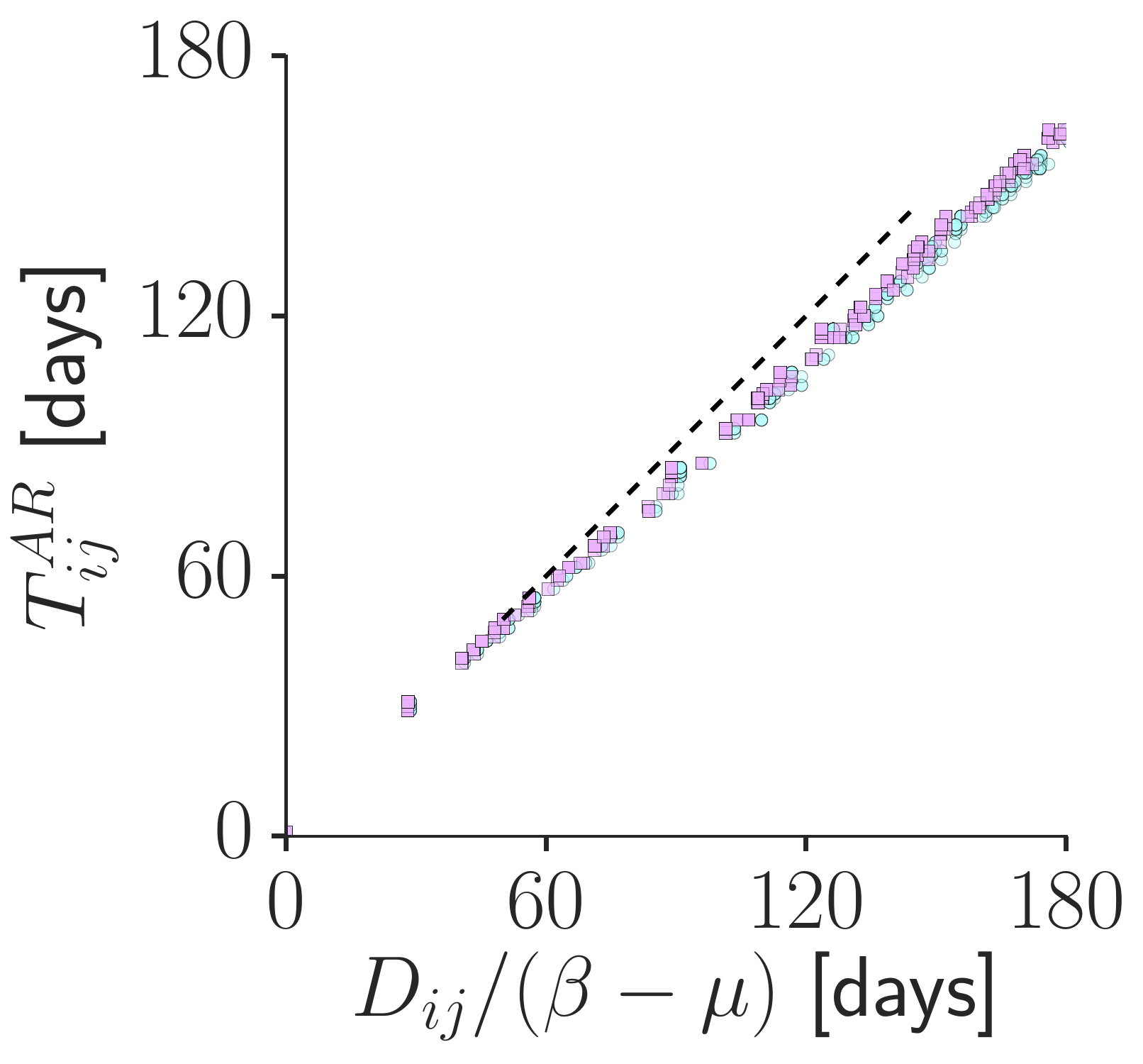}}
\qquad
\subfigure[]{\includegraphics[scale=0.18]{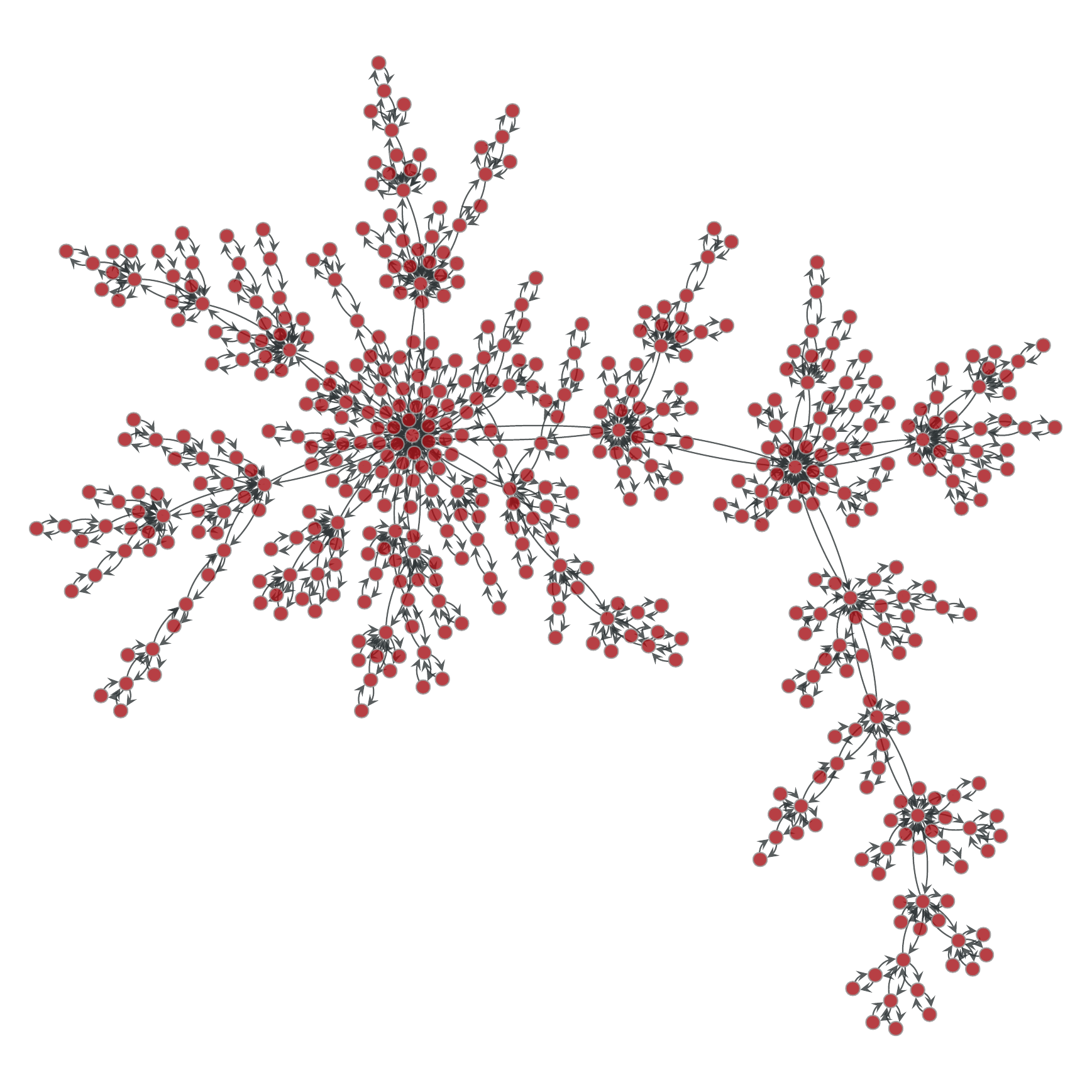}}
\qquad
\subfigure[]{\includegraphics[scale=0.28]{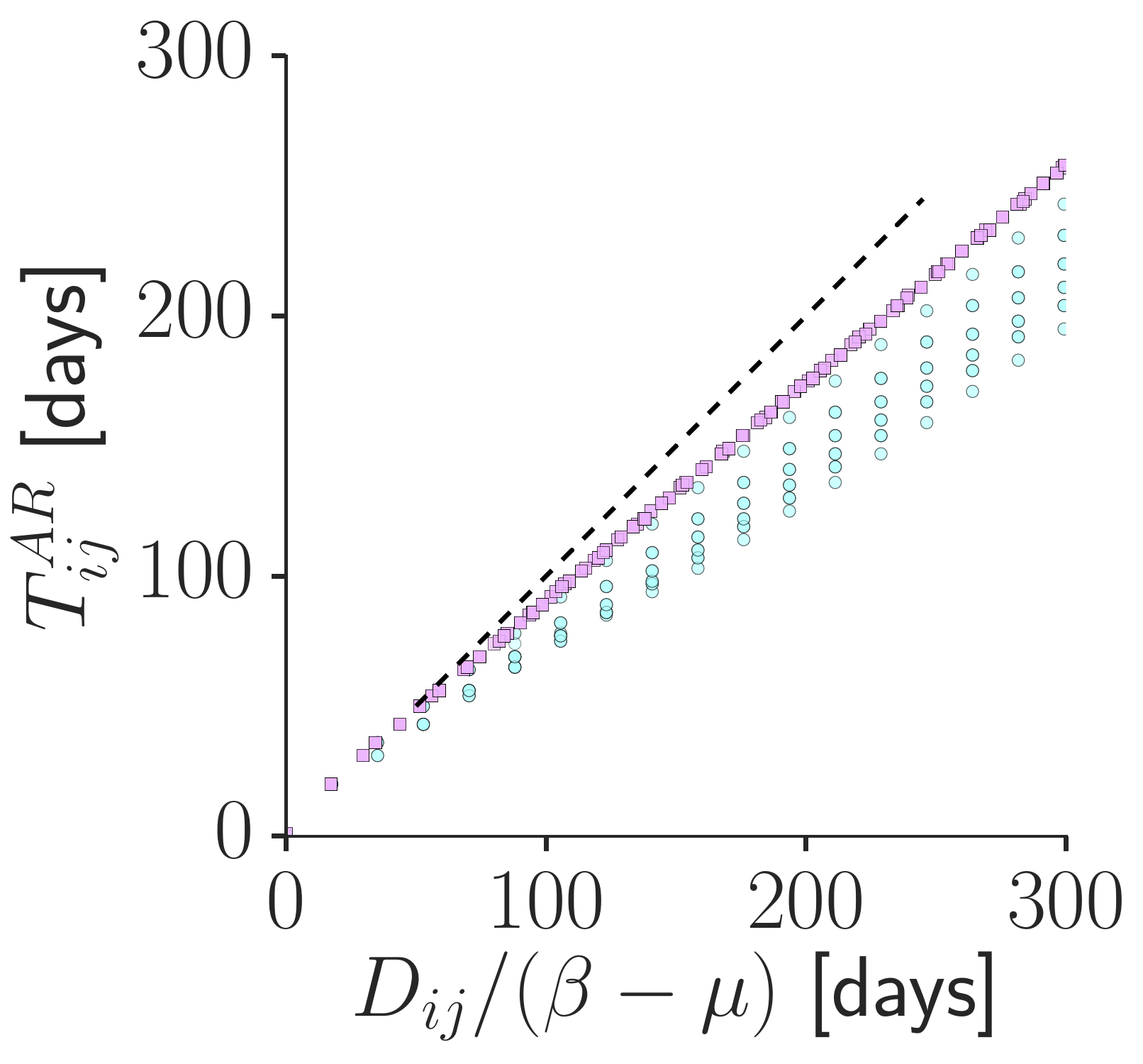}}
\qquad
\subfigure[]{\includegraphics[scale=0.18]{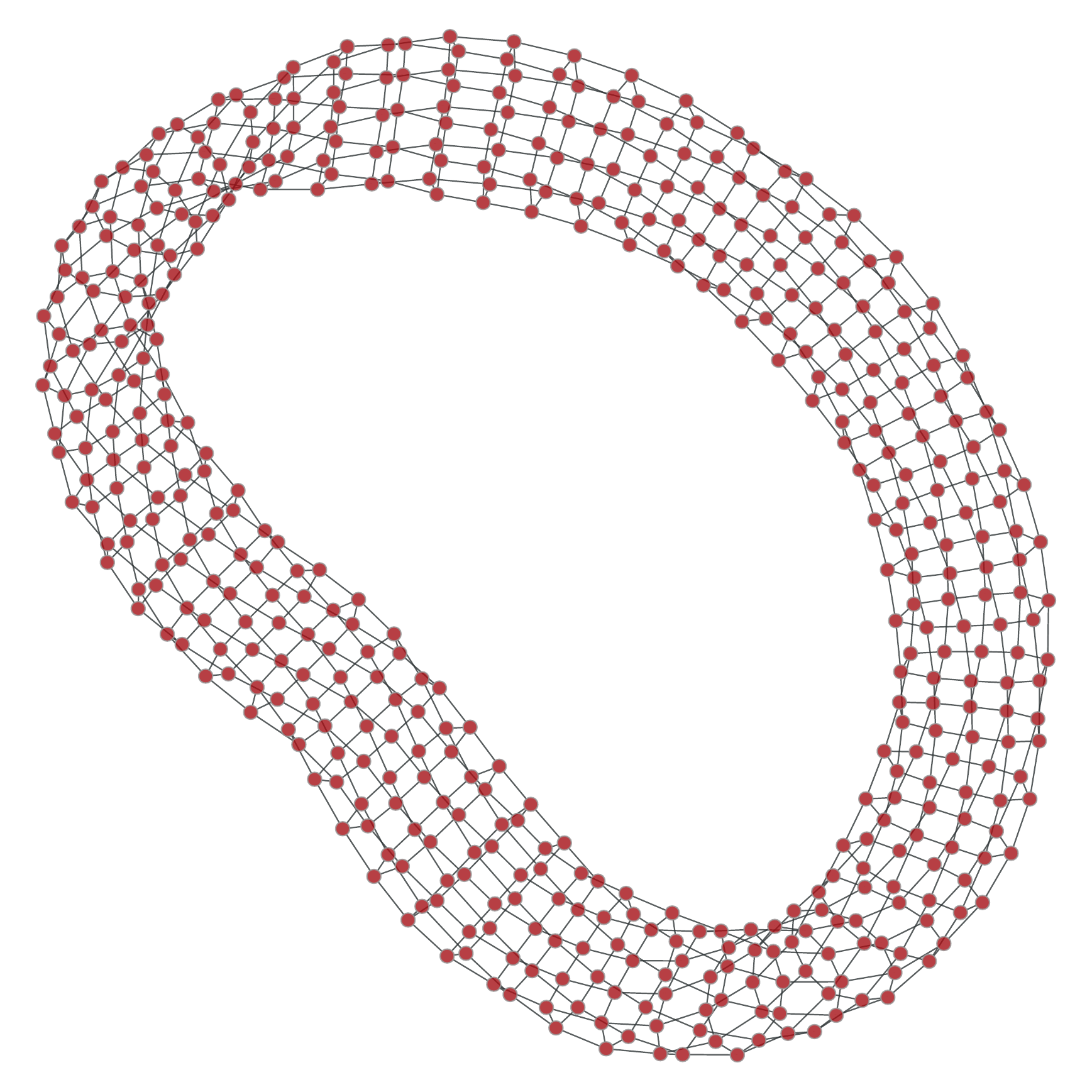}}
\caption{\label{corr_bara} (Color online) Correlation of the shortest-path (light-blue) and random-walk (violet) approach with the simulations arrival time for an unweighted  Barab\'asi-Albert network (a) and a two-dimensional lattice embedding (c) both consisting of $V=500$ nodes. In (b) and (d) the corresponding networks visualisation. The number of edges is respectively $E=998$  and $E=1000$. Parameters as in Fig~\ref{corr1}.} 
\end{figure*}

For highly heterogeneous topologies, such as the air-traffic network \cite{barrat2004architecture}, only a small number of paths contributes to  ${D}^{\text{MP}}_{ij}$. Taking the limit of the dominant contribution in \eqref{effrw1}, which corresponds to selecting the dominant path in \eqref{effchi1},  allows one to neglect the sum over the walks (paths) and it yields as for \eqref{guiz}
\begin{equation}
\widetilde{D^{\text{RW}}_{ij}}(\delta)  =   -\ln \left( e^{-\widetilde{n}\delta} H_{ij}(\widetilde{n})\right) = D^{\text{SP}}_{ij}(\delta).
\end{equation}
In Fig.~\ref{corr_usa} the comparison between the shortest-path and random-walk approach for the USA air-traffic network \cite{colizza07} shows that the results presented here are robust also for Erd\H{o}s-R\'{e}nyi networks. The  correlation coefficients are respectively $R^2_{\text{SP}} = 0.99$ and $R^2_{\text{RW}} = 1.00$ for the USA air-traffic network and  $R^2_{\text{SP}} = 0.94$ and $R^2_{\text{RW}} = 1.00$ for the randomized Erd\H{o}s-R\'{e}nyi network. An higher correlation and stability of the random-walk approach is also observed in the case of artificial networks, as for unweighted  Barab\'asi-Albert \cite{Barabasi509} and lattice models, see Fig.~\ref{corr_bara}. The correlation coefficients for the latter are $R^2_{\text{SP}} = 1.00$ and $R^2_{\text{RW}} = 1.00$ for the Barab\'asi-Albert network and  $R^2_{\text{SP}} = 0.99$ and $R^2_{\text{RW}} = 1.00$ for the lattice.

\section{Conclusions}

In summary we have presented a generalization of the concept of effective distance by overcoming the restriction of simple path propagation of a disease. The proposed random-walk effective distance includes the previously defined shortest-path measure as a particular case. 
The remarkable correlation found with the infection arrival time can be explained as follows. The contribution of looped trajectories  in the propagation of physical information is neglected thanks to  the decreasing exponential in the walk length. The latter serves as damping such contributions for long walks, and in particular allows to neglect infinite loops contributions. 
In scenarios where multiple parallel paths are important, for instance in Erd\H{o}s-R\'{e}nyi graphs or regular lattices, the assumption of a single dominant path breaks down and the measure proposed here can be used as an efficient alternative.
The predictive power of the random-walk effective distance can be  used for containment strategies and estimation of arrival times for real  global pandemics from the underlying networks topology.
The random-walk metric can in fact  be generally applied to any weighted and directed network besides the transportation ones, for instance in the context of social interactions and rumour spreading. 
For unweighted locally tree-like networks both the shortest-path and random-walk effective distances  yield maximum correlation with the simulated arrival time, as the shortest path tends to dominate.

From a theoretical point of view our results show that the average infection arrival time in  a meta-population model can be approximated by the cumulant generating function of the hitting time for a Markov chain. In fact, the generating function approach can also be used to formally derive the latter from the first moments of the Gumbel distribution \cite{gautreau2007arrival}.
The connection with the cumulant generating function allows for an interpretation within statistical physics. In particular this would explain how the different approaches are connected in terms of the entropy associated to paths of fixed length \cite{Kivimaki2014600} \cite{Bavaud2012}.
This observation links disease spreading on complex networks with  a generic diffusion process.
Further developments and extensions of our  results include the generalization to temporal networks by considering a set of transition matrices, one for each time step \cite{lentz2013unfolding,10.1371/journal.pone.0151209}.

\section*{Acknowledgements}

We thank T. Isele for insightful discussions and technical assistance.
This work was developed within the scope of the IRTG 1740 / TRP 2015/50122-0, and funded by the DFG / FAPESP. A. K. and P. H. acknowledge support by Deutsche Forschungsgemeinschaft in the framework of SFB910.

\bibliographystyle{apsrev4-1}
\bibliography{effective_distances}

\end{document}